# How can a geothermal storage system be optimally integrated into a local district? A case study.

U Schilt[1,2*], S Vijayananda[1], S Schneeberger[1], M Meyer[1], S Iyyakkunnel[1], P M Vecsei[1], P Schuetz[1]

[1] Competence Centre for Thermal Energy Storage, School of Engineering and Architecture, Lucerne University of Applied Sciences, Horw, Switzerland
[2] School of Architecture, Civil and Environmental Engineering, École Polytechnique Fédérale de Lausanne (EPFL), Lausanne, Switzerland

* Corresponding author: ueli.schilt@hslu.ch

**Abstract**. Achieving net-zero targets requires the phase-out of fossil-based heating. A major challenge is the seasonal mismatch between renewable heat supply and demand. District heating networks often dispose of excess heat in summer and rely on fossil backups in winter. Large-scale thermal energy storage offers a solution by storing surplus summer heat for use during winter, thus reducing the need for fossil fuels. This study investigates the feasibility of a large-scale thermal storage system at a power production site that supplies a large district heating network in the city of Bern, Switzerland. Specifically, the study examines the potential of a geothermal storage system to offset fossil fuel heat generation in winter by utilising heat stored during the summer months. Using a Python-based multi-energy system model, we simulate the optimal operation of the geothermal storage system with respect to cost and emissions, considering both supply and demand on an hourly basis over one year. Multi-objective optimisation is applied to generate a Pareto-optimal front. The results show that the geothermal storage system eliminates the requirement of 8 GWh of gas-powered heat supply and increases the waste heat utilisation by 20%, therefore lowering emissions. This effect is further increased when combined with an expansion of the district heating network, as individual, emission-heavy heaters are replaced by low-emission heat from the district heating network. The findings presented in this study can prove useful when evaluating similar systems across Switzerland.

## 1. Introduction

In Switzerland and across Europe, district heating networks (DHN) are expanding to meet climate targets and reduce reliance on fossil fuels. Switzerland's energy strategy for net-zero greenhouse gas emissions by 2050 places strong emphasis on decarbonizing space heating. District heating – the distribution of heat from centralised renewable or waste sources – is considered a key solution to achieve this [1]. District heating offers several advantages over individual heating: heat can be generated at a larger scale, therefore increasing efficiency and flexibility, and it can integrate diverse renewable sources and recover waste heat (e.g. from industry or waste incineration) that would otherwise be lost [2], [3]. In Switzerland, waste-to-energy plants already supply a large share of DHN heat [4]. Integrating a thermal energy storage (TES) system into a DHN can further improve its performance as it allows surplus heat from summer to be stored and used in winter. This provides valuable seasonal balancing by shifting energy from low-demand periods to high-demand periods, making greater use of renewables and reducing peak loads. Simulation studies show that adding a TES can raise annual waste heat utilisation, reduce $CO_2$ emissions, and improve efficiency [5], [6], [7].

In the city of Bern, Switzerland, the energy utility company Energie Wasser Bern (EWB) has been operating a DHN since 1960. The majority of the heat supplied to the DHN is generated at the power production site (PPS) Energiezentrale Forsthaus located on the city outskirts of Bern [8]. To increase the share of renewable heat supply, EWB has been expanding its DHN since 2020. There have also been plans to integrate a geothermal energy storage system with a capacity of 12-15 GWh. The aim was to capture surplus heat generated by the waste-to-energy plant during summer and store it for use in winter, when the heating demand is high. However, after a detailed geological analysis, it was concluded that the planned geothermal storage could not be built due to unsuitable subsurface conditions. Nonetheless, the same or similar energy system configurations could prove relevant for the decarbonisation of the energy supply in other locations where the boundary conditions are suitable for the implementation of a large-scale TES system. Therefore, in this study, we use the PPS Energiezentrale Forsthaus as an exemplary case to investigate the role of a large-scale TES system in a DHN. We model the heat and electricity demand of the city of Bern and carry out an optimisation study that simulates the optimal supply of a DHN covering a specified share of the city's heat demand. We analyse three scenarios: (1) optimal supply of the current DHN without TES, (2) optimal supply with TES, and (3) optimal supply of an expanded DHN with TES. The study focuses on the reduction of emissions but also considers the trade-off between emissions and system operation costs.

## 2. Methodology

In this study we deploy the District Energy Model (DEM), a model recently developed in-house for the analysis and optimisation of mixed-use multi-energy systems in Swiss communities ranging from a few buildings to regional areas. The model focuses on balancing local energy demand and supply. An earlier version of the model was presented in a case study for a Swiss municipality [9]. DEM is a linear model with hourly resolution implemented in Python. An integrated optimisation module, based on the mixed-integer linear programming framework Calliope [10], allows the analysis of energy systems with a specified optimisation objective (e.g. cost or emissions minimisation). In addition, DEM can perform multi-objective optimisation using the epsilon-constraint method [11]. Optimisation is carried out using the Gurobi solver [12]. DEM relies on public data for demand calculations and automated parameterisation. Information on individual buildings, including the type of installed heating system, is taken from the Federal Register of Buildings and Dwellings (RBD) [13]. The space heating demand of each building is modelled based on building characteristics and ambient weather data [14], [15]. Domestic hot water demand is modelled using values from the SIA standard [16]. The residential electricity demand is also modelled bottom-up for each building based on building characteristics. The demand is then scaled to standardised demand profiles taken from Rinaldi et al. [17]. Industrial electricity load profiles are modelled using a top-down approach based on data from the national transmission grid operator, Swissgrid. More detailed information about the modelling approach of DEM can be found in preceding work [9], [15], [18], [19], [20].

### 2.1. Power production site Energiezentrale Forsthaus

This study focuses on the optimal operation of the power production site (PPS) Energiezentrale Forsthaus. The PPS is represented in the model using the layout displayed in Figure 1. There are three sources of energy. The waste-to-energy plant generates both heat and electricity, acting as a Combined Heat and Power system (WtE CHP). The heat can be used directly by the DHN. The gas turbine generates electricity and high-temperature steam. The steam can be utilised in a steam turbine to generate both electricity and heat for the DHN. The coupled system of gas and steam turbines constitutes the Combined Cycle Gas Turbine (CCGT) system. A wood-fired plant also produces steam, which is converted into electricity and heat using the same steam turbine as the CCGT system. Together, the wood-fired plant and the steam turbine form the Wood-fired Combined Heat and Power plant (Wood CHP). The parameter values for the technologies relevant to the DHN are provided in Table 1. We constrain the WtE CHP plant to operate at a minimum of 50% capacity per time step to ensure continuous waste processing, independent of energy optimisation goals. The other technologies are not subject to part-load constraint.

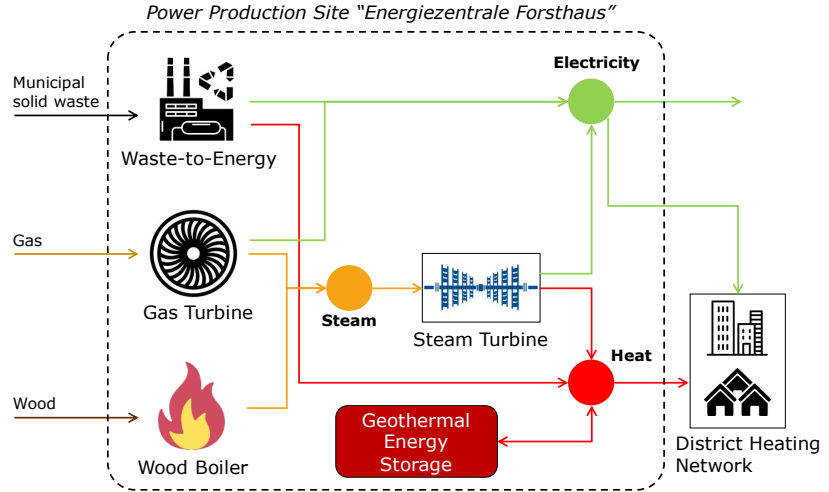

**Figure 1.** Energy system layout used for modelling the power production site Energiezentrale Forsthaus. The model uses some simplifications compared to the real system (an intermediate waste heat boiler between gas turbine and steam turbine, as well as a gas boiler for peak loads are omitted). Heat is supplied to the DHN and electricity is supplied to the city of Bern, including the DHN.

**Table 1.** Parameter values for energy technologies connected to the district heating network: electric capacity ($P_{el}$), electric conversion efficiency ($\eta_{el}$), thermal conversion efficiency ($\eta_{th}$), $CO_2$-equivalent intensity ($I_{CO2,eq}$), available fuel ($M_{fuel}$), and fuel costs ($c_{fuel}$).

| Technology | $P_{el}$ | $\eta_{el}$ | $\eta_{th}$ | $I_{CO2,eq}$ | $M_{fuel}$ | $c_{fuel}$ | Sources |
|---|---|---|---|---|---|---|---|
| WtE CHP | 16 MW | 0.20 | 0.45 | 775 g/kWh$_{el}$ | 110'000 t/a (msw[1]) | -30 Rp/kg | [21], [22], [23] |
| Gas turbine | 46 MW | 0.35 | 0.53 | 760 g/kWh$_{el}$ | inf. (natural gas) | 10 Rp/kWh | [21], [24], [25] |
| Wood boiler | - | - | 0.86 | 27 g/kWh$_{th}$ | 12'000 t/a (wood) | 24 Rp/kg | [21], [22], [26] |
| Steam turbine | 27 MW | 0.07 | 0.70 | - | - | - | [21], [24] |

[1]municipal solid waste

### 2.2. Modelling the currently installed district heating network

Estimating the current DHN share (i.e. the share of the city's heat demand covered by the DHN) is not straight-forward. The RBD database lists the heating system type for each building. Based on the DEM's heat demand model and information from the RBD, 13.1% of the heat demand in the city of Bern is supplied by district heat. However, a comparison of the information provided on the utility's website regarding districts connected to the DHN and the information provided in the RBD shows that a lot of the information contained in the RBD is outdated. For example, the district Looslistrasse-Untermattweg has been connected to a DHN as reported on the utility's website (Figure 2) [27]. However, in the RBD database, only two buildings in this district are listed as connected to the DHN, while other buildings are still mostly listed as using oil or gas boilers for heating. In this study, we increase the share of the demand supplied by DHN from 13.1% to 18% to represent the current system. This is to account for demand that is already supplied by DHN but not yet reflected in the RBD.

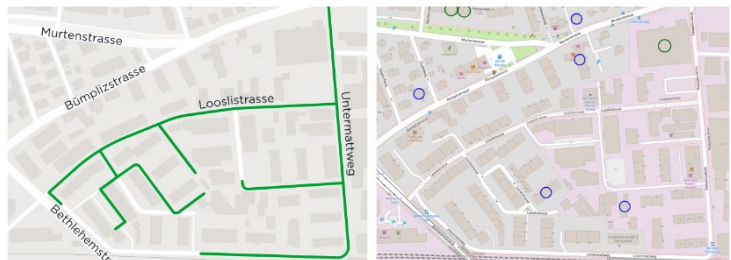

**Figure 2.** The issue of outdated data in the RBD demonstrated using the example of the district "Looslistrasse-Untermattweg". The DHN is reported as completed by EWB (left) [27]. However, the data in the RBD does not yet reflect all DHN connections (right; connected buildings are indicated with blue circles) [28].

*2.3. Scenarios*

We investigate three scenarios for the energy supply of the city of Bern and optimise the operation of the PPS for one year. The decentralised heat supply to buildings not connected to the DHN is also simulated but not optimised. This means the infrastructure (i.e. heating systems and sources) remains as currently installed. In each scenario the heat supply to the DHN is optimised both for costs and emissions reduction. In addition to single-objective optimisations (cost and emissions), a Pareto-optimal front is generated for each scenario by performing multi-objective optimisation based on the epsilon-constraint method. Scenario 1 represents the energy system infrastructure with a DHN share of 18% (system shown in Figure 1 without the TES) that is currently installed. The operation of the DHN-supplying technologies (WtE CHP, CCGT, Wood CHP) is optimised. Scenario 2 builds on Scenario 1. In addition, a large-scale thermal energy storage (TES) system of 15 GWh capacity is integrated in the DHN (system shown in Figure 1). Scenario 3 builds on Scenario 2. In addition, an expansion of the DHN share up to 50% of the city's heat demand is considered. In this scenario, an increased capacity of the Wood CHP is also permitted.

## 3. Results

In Scenario 1, 18% of the city's heat demand is supplied by the DHN. The remaining share is supplied by decentralised heating technologies, namely oil boilers, gas boilers, wood boilers, electric heaters, and heat pumps (Figure 4 top). The results show that a large share of the city's heat demand is still met with fossil fuels (oil and gas), while for example, the share of the demand supplied by heat pumps remains relatively small. The extent to which this is due to outdated RBD data is unclear. In Figure 4 (top), which shows the daily heat supply split by technologies, the supply of the DHN (CCGT, WtE CHP, Wood CHP) is optimised for emissions reduction. It can be observed that during summer the WtE CHP plant generates surplus heat that is not used. Annually, the largest amount of heat is supplied to the DHN by the WtE CHP plant (97 GWh), followed by the Wood CHP plant (84 GWh), and the CCGT contributing a comparatively small amount (8 GWh) (Figure 5, Scenario 1). The CCGT only supplies heat during the heating season, when heat from the WtE CHP and Wood CHP plants is not sufficient. In Scenario 2, with the integration of a TES, some of the surplus heat generated by the WtE CHP plant during summer can be stored and used during the heating season (Figure 4 bottom), thus reducing emissions through lower fuel consumption. The WtE CHP plant still supplies the largest share of heat to the DHN, increasing to 116 GWh (+20%). However, not all surplus heat can be utilised, suggesting that the TES capacity could still be increased. The Wood CHP plant provides 72 GWh (-14%), while the CCGT is no longer required in this scenario, therefore reducing fossil heating by 8 GWh (Figure 5, Scenario 2). In Scenario 3, due to the expansion of the DHN to 50%, the heat supply from both the WtE CHP plant and the Wood CHP plant increases to 145 GWh and 378 GWh, respectively (Figure 5, Scenario 3).

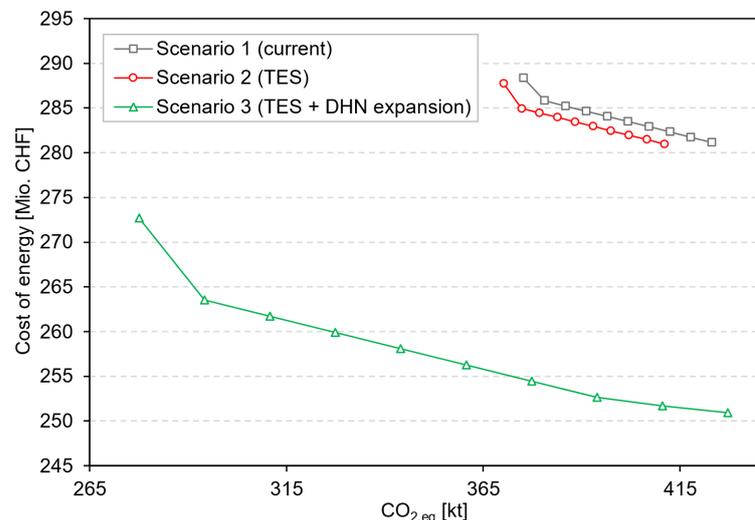

**Figure 3.** Pareto optimal front for Scenario 1 (18% district heating network (DHN) demand share), Scenario 2 (18% DHN demand share and thermal energy storage (TES)), and Scenario 3 (DHN expansion to 50% demand share and TES) generated based on the epsilon-constraint method.

The multi-objective optimisation shows that a reduction in emissions comes at an increased operational cost for each scenario. Figure 3 shows the Pareto-optimal front for each scenario. The integration of a TES into the DHN (Scenario 2) reduces the cost of operation for a given emissions limit. In other words, for the same operational cost, heat can be supplied with lower emissions. The same effect is observed when expanding the DHN up to 50%, but at a much larger scale.

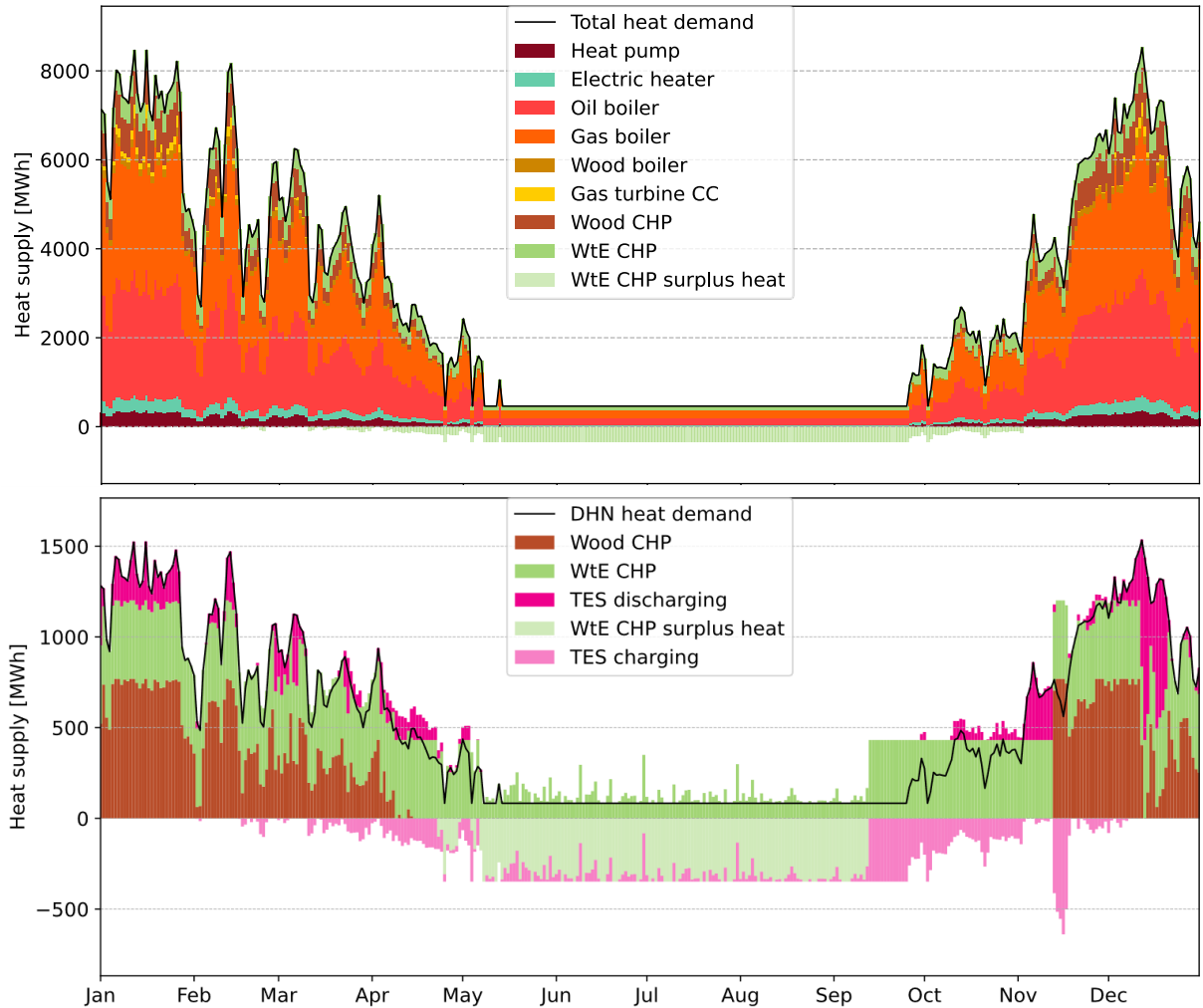

**Figure 4.** Daily heat supply split by supply technologies and optimised for emissions reduction. Top: Heat supply to the city of Bern assuming an 18% coverage by DHN supplied by the power production site (PPS) (Scenario 1). Bottom: Heat supply from the PPS Energiezentrale Forsthaus to the district heating network with operation optimised for emissions reduction und use of geothermal storage (Scenario 2). Some of the surplus heat generated by the waste-to-energy plant can be shifted for use during the heating season using the TES.

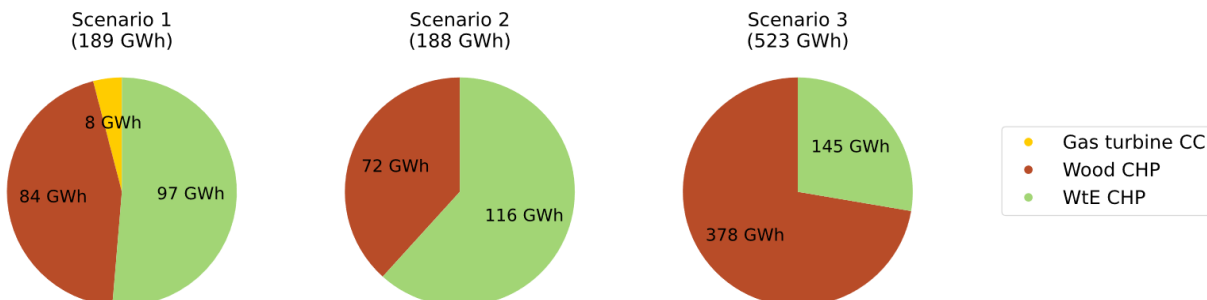

**Figure 5.** Heat supplied to the district heating network (DHN) for Scenario 1 (current), Scenario 2 (TES), and Scenario 3 (TES + DHN expansion) in the case of operation optimised for emissions reduction.

## 4. Discussion and conclusions

Our model results show that the integration of a large-scale thermal energy storage (TES) system into a district heating network (DHN) can reduce emissions and operational costs, even without replacing or increasing the capacity of existing supply technologies. This effect is particularly pronounced when a technology generates surplus heat during the low-demand season (i.e., summer), as is the case for the waste-to-energy combined heat and power plant. The results also indicate that, despite TES integration, surplus heat remains, suggesting that an increase in TES capacity could lead to further emission reductions. It should be noted that this study focuses exclusively on operational costs. Therefore, when considering the integration of a TES into a DHN, operational savings must be weighed against the associated capital investment. The study also demonstrates that expanding the DHN can reduce emissions at a large scale. This is likely because heat demand not covered by the DHN is predominantly met by individual, fossil-based, emission-intensive heating systems (e.g., oil and gas boilers). Expansion of the DHN allows these systems to be replaced with lower-emission district heating. However, this effect would be less pronounced - or potentially absent - if the individually supplied buildings were instead equipped with low-emission systems such as heat pumps powered by renewable electricity.

Limitations to this study include the uncertainty associated with publicly available data sources used in the modelling (e.g., the RBD), as well as the sensitivity of the results to input assumptions such as cost and emission factors. For instance, fluctuations in the price of natural gas could lead to substantially different outcomes. These effects should be further explored in future work using sensitivity analysis.

Overall, these results demonstrate the potential of TES and DHN expansion as effective strategies for reducing emissions in urban energy systems and that both approaches can make a meaningful contribution to the decarbonisation of the heating sector. Moreover, the developed model is transferable and can thus be applied to other DHN systems with similar layouts and characteristics.

**Acknowledgements**

The research published in this publication was carried out with the support of the Swiss Federal Office of Energy as part of the SWEET consortium EDGE. The authors bear sole responsibility for the conclusions and the results presented in this publication.